\documentclass[11pt, oneside]{article}
\usepackage{geometry}
\usepackage{graphicx}	
\usepackage{amssymb}
\usepackage{xcolor}
\usepackage{siunitx}
\usepackage{multirow}

\title{Surface Modifications of PCB-Based Plasma Sources Induced by Atmospheric Plasma: A Comparative Study of Dielectric and Electrode Materials}
\author{Jonathan Gail, Alisa Schmidt, Markus H. Thoma}

\begin{document}
\maketitle
\hrule
\section*{Abstract}
The study investigated the effects of atmospheric plasma on various dielectric materials (FR-4, PTFE, Al\(_2\)O\(_3\)) and electrode materials (copper, silver, gold-plated copper) used in surface micro-discharge plasma sources. XPS and laser microscopy were used to analyze changes in surface properties and chemical composition after 10 hours of plasma exposure.\\
Al\(_2\)O\(_3\) showed the highest resistance, with no significant changes in Al/O ratio or oxidation state. PTFE underwent oxidation, with fluorine substituted by oxygen, forming carbonyl, hydroxyl and aldehyde groups. FR-4 showed the most substantial changes, with etching of the epoxy matrix exposing glass fibers. This was in line with the XPS results, which show a higher O/C-ratio, less epoxy groups and higher nitrogen signals. Silver electrodes were most resilient, with sintered particles redistributed across the Al\(_2\)O\(_3\) surface. Copper and gold-plated copper electrodes were more susceptible to oxidation and degradation, especially in areas exposed to plasma filaments.\\
The results highlight the importance of carefully selecting materials based on resistance to plasma exposure and compatibility with the application.\\
\hrule\vspace{4mm}
\section{Introduction}
Atmospheric pressure plasma sources have gained significant attention in recent years due to their wide range of applications in surface modification, sterilization, and materials processing \cite{Hsiao,mi14071331,9627078}. The etching capability of atmospheric plasma is a well-known effect and subject of current research \cite{Hsiao,https://doi.org/10.1002/app.54449}. Additionally, plasma-induced surface modifications such as changes in contact angles and chemical composition of different materials have been investigated \cite{9627078}. However, while much research has focused on the effects of plasma on target materials, less attention has been paid to potential changes in the dielectric and electrode materials of the plasma sources themselves. As other materials are subject to plasma-induced changes, it is likely that the components of plasma sources are also affected, potentially even more rapidly.\\
Previous research has shown that electrode geometry can significantly influence plasma chemistry \cite{Joni}. This suggests that geometry may also play a role in the surface modification of plasma sources. To investigate this, we examined three plasma source concepts with different electrode geometries, each exhibiting distinct discharge behaviors and filament patterns.
The choice of dielectric material is crucial for the performance and longevity of plasma sources. Common materials used in plasma research include FR-4, polytetrafluoroethylene (PTFE), and aluminum oxide (Al2O3) \cite{pr10010104,CEPLANT2002,bilek2020plasma}. Each of these materials offers unique properties in terms of chemical resistance, dielectric constant, and processability.\\
In this study, it was to investigate the effects of atmospheric plasma on three dielectric materials (FR-4, PTFE, Al2O3) and three electrode materials (copper, silver, gold-plated copper) used in surface micro-discharge plasma sources. By utilizing X-ray photoelectron spectroscopy (XPS) and laser microscopy, the changes in surface properties and chemical composition after extended plasma exposure were analyzed. This enables conclusions on the length of the polymers chains and the oxidation-states of the material, which is directly linked to the plasma resistance of the plasma source. Hence, this approach allows the evaluation of the suitability of different materials for long-term use in plasma sources and provide insights into optimizing their design for specific applications.

\section{Experimental Setup}
The plasma sources were based on the results of further research articles \cite{Joni,CEPLANT2011}. The sources were operated for 10 hours within a closed volume of one liter. During this, the ozone (O\(_3\)) concentration was measured by absorption spectroscopy at \SI{254}{nm}. The supply voltage parameters were tuned to a constant ozone production rate that reached a saturation concentration of \SI{300}{ppm}. This ensured a comparable capability for chemical changes between the plasma sources. The following chapter emphasizes the differences between the sources.

\subsection{Plasma Sources}
\begin{figure}[ht!]
   \centering
   \includegraphics{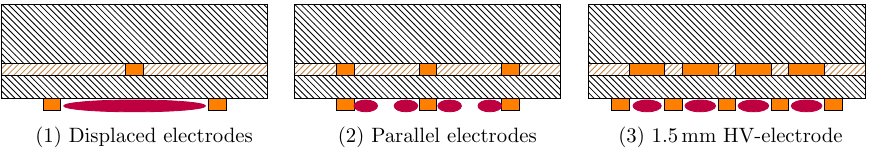}
   \caption{Three used dielectric- and electrode geometries, red: plasma, orange: copper electrode
(lower electrode is grounded), black: dielectric, brown: epoxy resin}
   \label{fig:Geo}
\end{figure}

To complement previous studies, three different electrode geometries were analyzed to evaluate their effects on the plasma chemistry and their changes during plasma operation \cite{Joni}. The electrode geometries, which were the result of an optimization process, are shown in figure \ref{fig:Geo}. The first setup had displaced electrodes with a distance of 3 mm. The second setup had parallel electrodes with a small plasma layer at the edges of the ground electrodes and the last one had displaced electrodes, with no distance and larger high voltage electrodes. This distributes the plasma-layer over the whole surface.
Additionally, three different dielectric materials were used to study the plasma effect on them. Those materials were commonly used in plasma research \cite{CEPLANT2002, LowTPlasma}. FR-4 is a substrate for printed circuit boards (PCB). The other dielectric material was PTFE. It is very inert concerning chemical stress and also available as a specialized form for PCBs. Yet, this form contains other components to make the substrate laminatable. The last dielectric used, is aluminum oxide (Al\(_2\)O\(_3\)) it also has the highest chemical resistance, but a higher dielectric constant.\\
To enhance the durability and reduce the formation of toxic products, silver sintered and gold plated copper electrodes were compared to standard copper electrodes.\\
The FR-4 substrate had copper electrodes, the PTFE sources were manufactured with gold plated copper electrodes and the conductive silver lines were sintered onto the aluminum oxide. For this process silver and glass powder were mixed (19 : 1) with small amounts of polyvinylalcohole (PVA) and mask printed onto the aluminium oxid. The sinter process was conducted at \SI{900}{\degree C} with \SI{100}{\degree C/h} slope and \SI{1}{h} at \SI{900}{\degree C}.
\subsection{XPS}
The plasma-induced changes of the dielectrics were measured by x-ray photoelectric spectroscopy (XPS). A comparison of pre and post plasma exposed samples showed the influence of the plasma on the molecular structures. Oxidizing, sputtering or bond breaking capabilities of the plasma were evaluated.\\
XPS can be used to quantitatively determine the chemical composition and oxidation state of components in a sample in a virtually non-destructive manner. The sample is bombarded with X-rays, causing photoelectrons to be emitted from the material. The difference between the known energy of the X-rays and the kinetic energy of the photoelectrons detected is the binding energy of the photoelectrons. As these are characteristic of a specific atomic orbital in a specific chemical environment, the origin of the photoelectrons can be determined.\\
Measurements were performed with a PHI 5000 VersaProbe II photoelectron spectrometer (Physical Electronics, Inc.) using monochromatic Al-K\(\alpha\) X-rays (1486.6 eV) with a beam diameter of \SI{200}{\mu m} at 50 W. The energy of the analyzer was set to 93.9 eV with a 0.80 eV step size for overview spectra and to 29.35 eV with a 0.25 eV step size for detail spectra. To counteract charges, electrons (approximately 1 eV) and argon ions (\(\le\) 10 eV) were neutralized during the measurement and all spectra were calibrated by setting the C 1s peak of the C-C / C-H bonds to 248.8 eV. The recorded spectra were analyzed using CasaXPS software (version 2.3.24PR1.0, Casa Software Ltd). The background was calculated using the Shirley method \cite{shirley1972high}.
\subsection{Laser Scanning Microscopy}
The surface changes of the plasma sources were evaluated by laser microscopy (KEYENCE VK-9700). This device measures the interference of a violet 408 nm laser signal from the surface with a reference path by apochromatic lenses (50x, 150x magnification), which gives information about the surface's profile. With this data the roughness is analyzed. Besides the different definitions of the roughness (e.g. mean, min-max, RMS, etc.), the arithmetic average and the root mean square average was used. The pre- and post plasma exposition values were compared to evaluate the physical changes of the surface.\\
The typography of the plasma sources was measured at the electrodes, at the filament formation point at the surface of the dielectric and the dielectric's surface in the middle between the electrodes.\\
Additionally to the 3D-data, the color information was mapped to the topography, which enabled the detection of different materials at the same height.
\section{Results}
The results from the XPS and laser microscopy were grouped by their dielectric materials, because there were significant effects measurable across the dielectrics. A cross-sectional analysis of the data, which were grouped by their electrode geometry, did not show uniform results for each geometry subset. Yet, each material specific result depended on the filamentary shape of the discharge, which decreased from plasma source (1) to (3).
\subsection{XPS}
\subsubsection*{Al\(_2\)O\(_3\)}
Based on its constituents, aluminium oxide should have an aluminium to oxygen ratio of 2:3, which was verified for the unused plasma source. However, this ratio was also be verified after plasma exposure. The variations of the measured values were in the range of \(\pm\) 1\%, which excludes a significant change. As no higher oxidation state of aluminium is known, this is in agreement with theoretical and experimental results, because those states were not even possible in an oxygen plasma environment \cite{Hsiao}. This is in line with the findings of our laser microscopy results.\\
After operating the plasma sources, more organic components were found on the surface. This can be attributed to the printed circuits, since small amounts of polyvinyl alcohol were used in the sintering process. Those could have been distributed over the surface by sputtering.\\
Silver was detectable within the substrate, because small amounts were sputtered and distributed over the surface. laser microscopy showed, that the deposition of the silver was filamentary shaped, like figure \ref{fig:Al2O3nachher} indicates, which supports the suggestion of an above mentioned sputter process.
\subsubsection*{PTFE}
Two effects were detected in every data set. First of all, fluorine became substituted by oxygen during the long-term plasma exposition. The oxygen signal O1s of C–O and C=O increased by 3 - 6\%. This is in line with the previous findings, that small amounts of oxygen were incorporated \cite{doi:10.1179/026708400101517161}. The amount of oxidized carbon increased from pre- to post-exposition up to a factor 1.5, which can be correlated to insertion of oxygen in the carbon chains and bond breakage. The oxidation results in the formation of carbonyl, hydroxyl and aldehyde groups.\\
These results suggest the evaluation of the changes in the length of the carbon-chains within the PTFE. Yet, the sputtering of PTFE and the different filament shapes lead to a non-homogeneously treated surface and the material was sputtered faster, than the bond breaks and oxidizing capabilities of the plasma can modify the material structure \cite{https://doi.org/10.1002/sia.740160186}.\\
 
\subsubsection*{FR-4}
\begin{figure}[ht!]
   \centering
\includegraphics[width=0.7\textwidth, height = 0.5\textwidth]{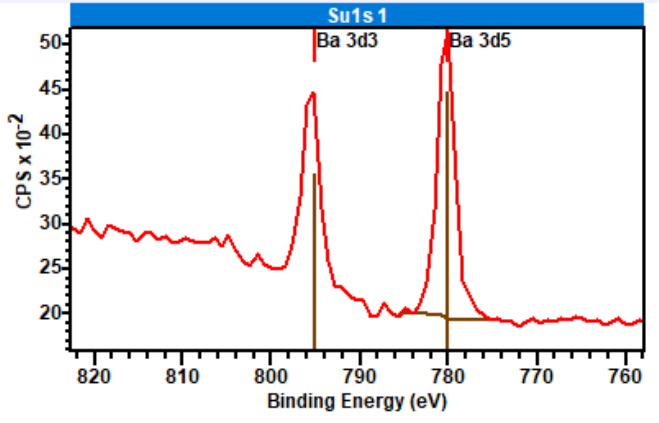}
   \caption{Barium lines in the post-exposure spectrum}
   \label{fig:Ba}
\end{figure}

The most significant results are measured with the FR-4 dielectric. The XPS data contained a strong silicon and barium signal after the plasma exposition. These peaks could refer to fiber-glass components in the resin based dielectric, which were uncovered by plasma etching process. Figure \ref{fig:Ba} shows the detailed spectrum, where the 3d\(_3/2\) and 3d\(_5/2\) peaks are clearly visible. This is correlatable with the results from the following laser microscopy section.
\begin{table}
  \centering
  \begin{tabular}{|l|c|c|c|c|}
  \hline
 & \bfseries pre  & \multicolumn{3}{c|}{ \bfseries post exposure}\\
  & \bfseries exposure & \bfseries(1) & \bfseries (2) & \bfseries(3) \\ \hline \hline
    \bfseries Ratio of O 1s/C 1s &\multirow{2}{*}{25\%} & \multirow{2}{*}{77\%} &\multirow{2}{*}{32\%} & \multirow{2}{*}{38\%}\\
  (C–O + C=O - SiO\(_2\)) / (C–C + C–H) &&&&\\\hline
  \bfseries Pre-to-post C 1s signal ratio &\multirow{2}{*}{100\%} & \multirow{2}{*}{11\%} &\multirow{2}{*}{22\%} & \multirow{2}{*}{58\%}\\
  Epoxy group (COC) & && &\\\hline
   \bfseries C 1s signal&\multirow{2}{*}{0\%} & \multirow{2}{*}{1.1\%} &\multirow{2}{*}{1.9\%} & \multirow{2}{*}{3.8\%}\\
   C–N & && &\\
\hline
\end{tabular} 
  \caption{XPS results from pre and post plasma exposition}\label{tab:XPSFR4}
\end{table}
Table \ref{tab:XPSFR4} shows the significant results from XPS analysis. The ratio of oxidized carbon atoms to C–C and C–H bonded carbon atoms increased for all plasma sources. In case of plasma source geometry (1) the increase reached a maximum of 308\% during plasma exposition. The oxygen signal is based on the O 1s from C-O and C=O, whilst the SiO\(_2\) component in this peak was subtracted. Due to many O 1s lines near the SiO\(_2\) and only one Si 2p peak, the subtracted oxygen signal was corrected by the doubled signal from Si 2p.\\
The dielectrics of the plasma sources showed a decreasing number of epoxy groups (COC) after plasma ignition, while more C atoms reach higher oxidized levels and oxygen was inserted into the molecules.\\
A nitrogen signal was measured for FR-4 after plasma exposition. It was matched to nitrate within a strong electron pulling environment. This underlined the weakness of this dielectric to chemical stress, because this could be caused by nitric acid etching. A typical reaction for this would be the formation by nitrate and water in a plasma environment
\cite{kojtari2013chemistry}. Another possibility for this signal would be the epoxy hardener, which contains silicon and nitrogen. Yet, none of those components were measurable before the plasma exposition.

\subsection{laser microscopy}
The results are displayed as microscopic images from pre- and post-exposure. Since the data of the surface roughness would exceed this page, the main effects are referenced and reported.
\subsubsection*{Al\(_2\)O\(_3\)}
\begin{figure}[ht!]
   \centering
   \includegraphics[width=0.3\textwidth, height = 0.4\textwidth]{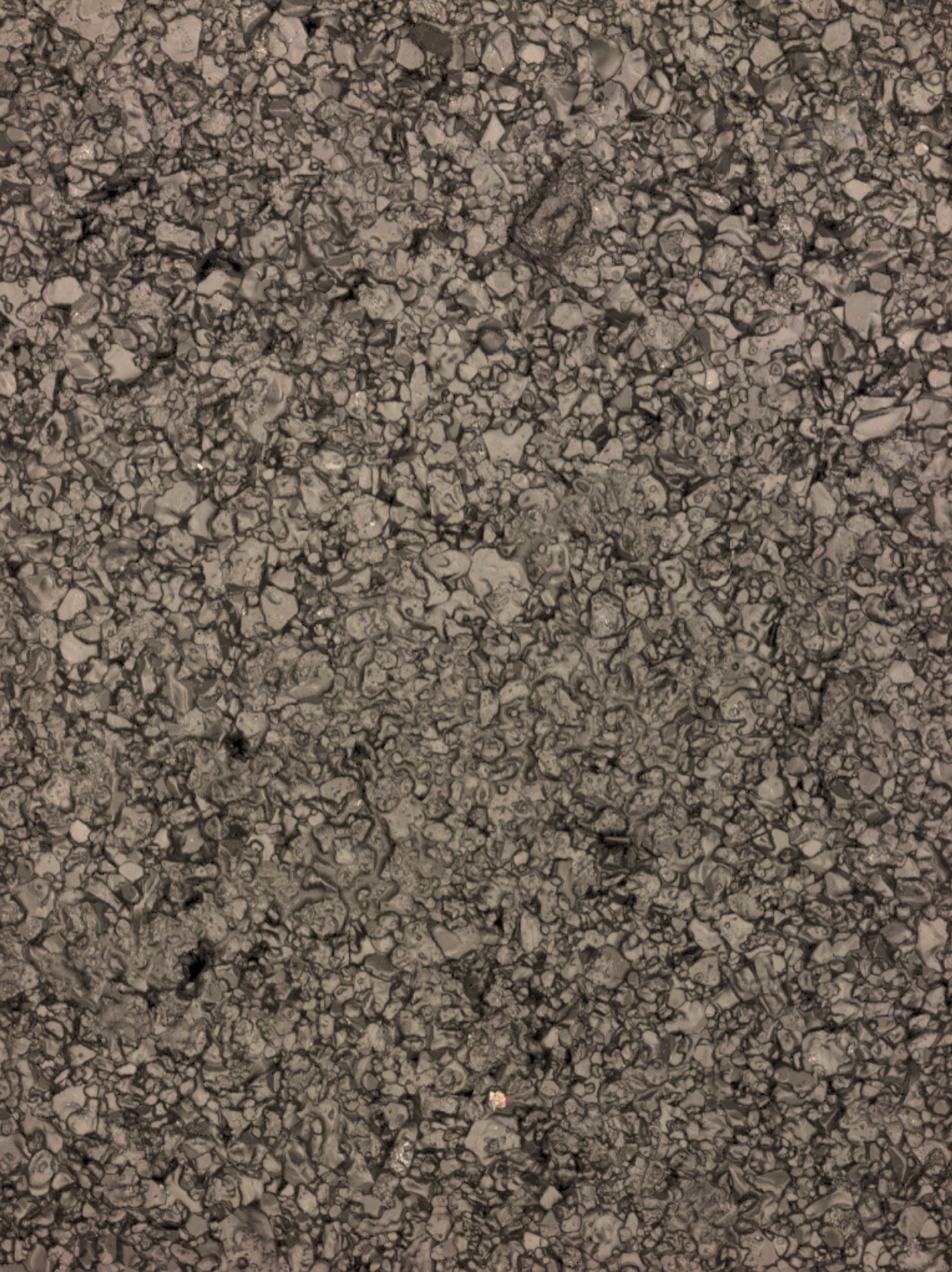}
   \includegraphics[width=0.3\textwidth, height = 0.4\textwidth]{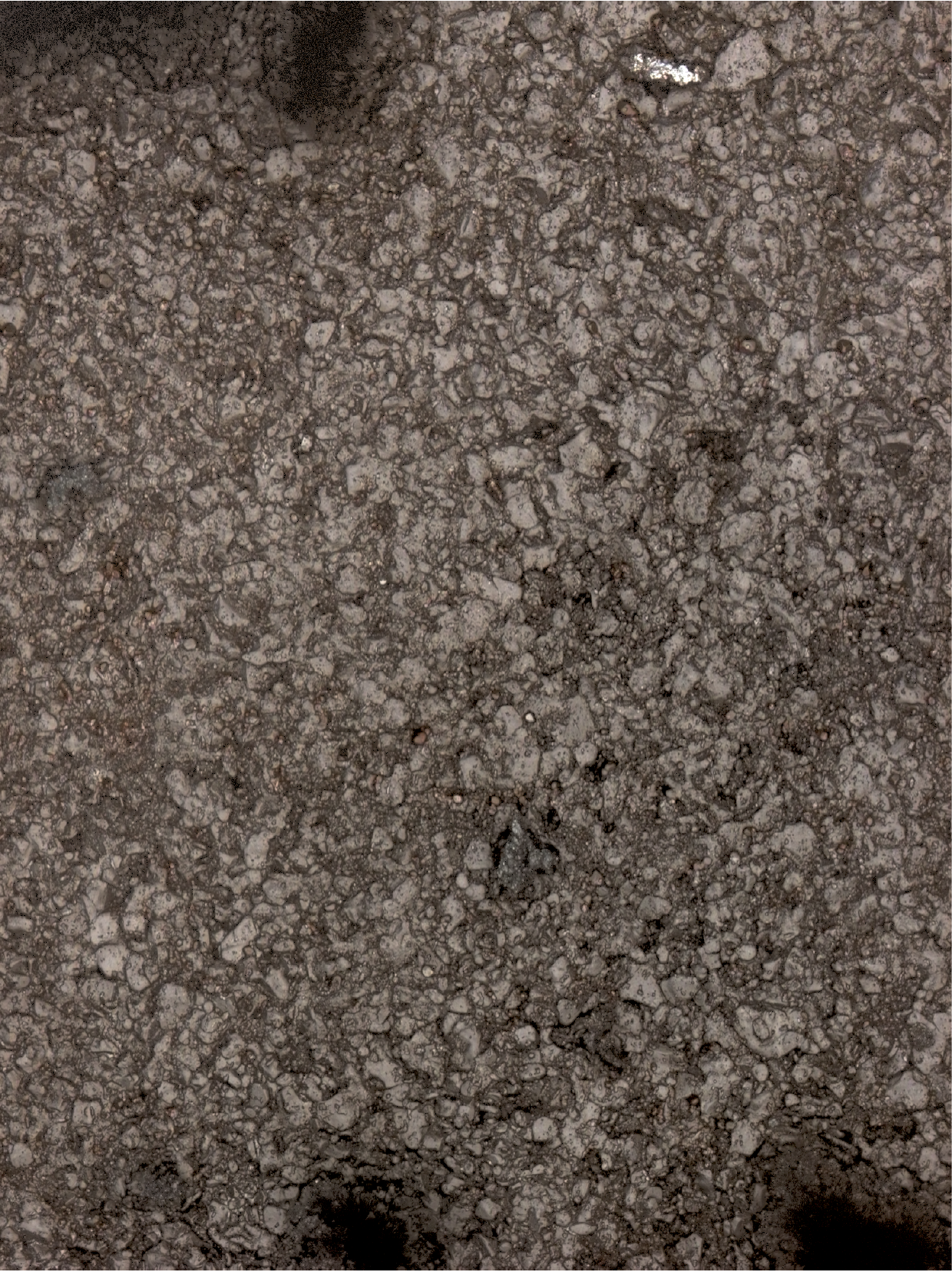}
    \includegraphics[width=0.3\textwidth, height = 0.4\textwidth]{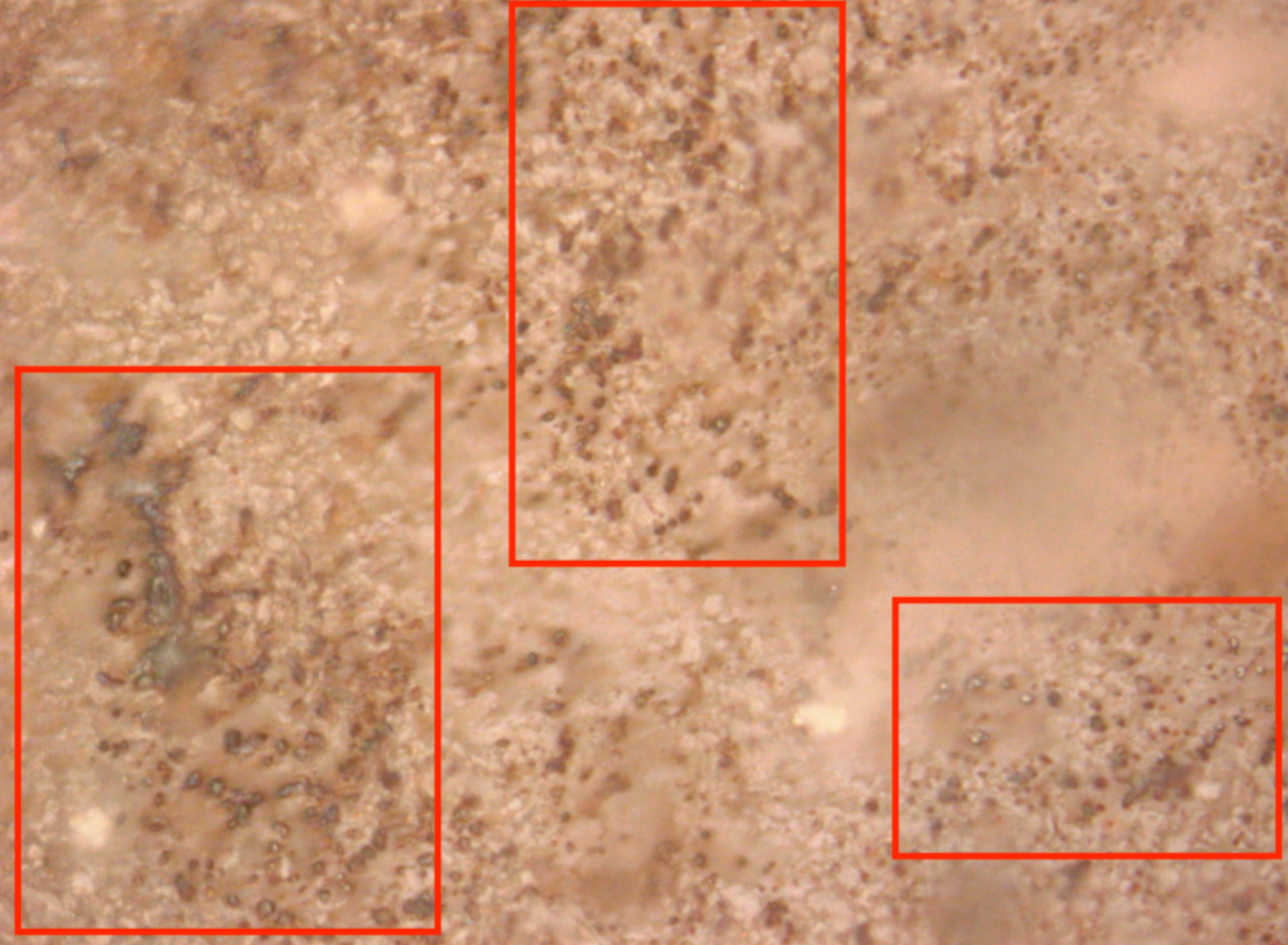}
   \caption{(a)Aluminiumoxide original; (b) Aluminiumoxide plasma exposed; (c) deposition of silver along the filaments}
   \label{fig:Al2O3nachher}
\end{figure}
Aluminiumoxide does not experience any significant change in surface roughness due to the plasma. Minor changes at the edges of the Al\(_2\)O\(_3\) particles in figure \ref{fig:Al2O3nachher} can be surmised. Yet, they are not measurable in the roughness. Small surface variations can be seen due to the deposition of silver material in the recesses of the surface structure in the right subfigure. This results in a slightly lower surface roughness. Figure \ref{fig:Al2O3nachher} shows the unchanged surface structure in the left image, the dielectric material after the exposition and the silver deposits in the recesses of the material in the right image. These silver deposits were observed locally in the areas of the filaments and decreased in intensity from plasma source (1) to (3), as does the effect of filamentary shaped structures on the dielectrics and the roughness.

\subsubsection*{PTFE}
\begin{figure}[ht!]
   \centering
   \includegraphics[width=0.4\textwidth, height = 0.4\textwidth]{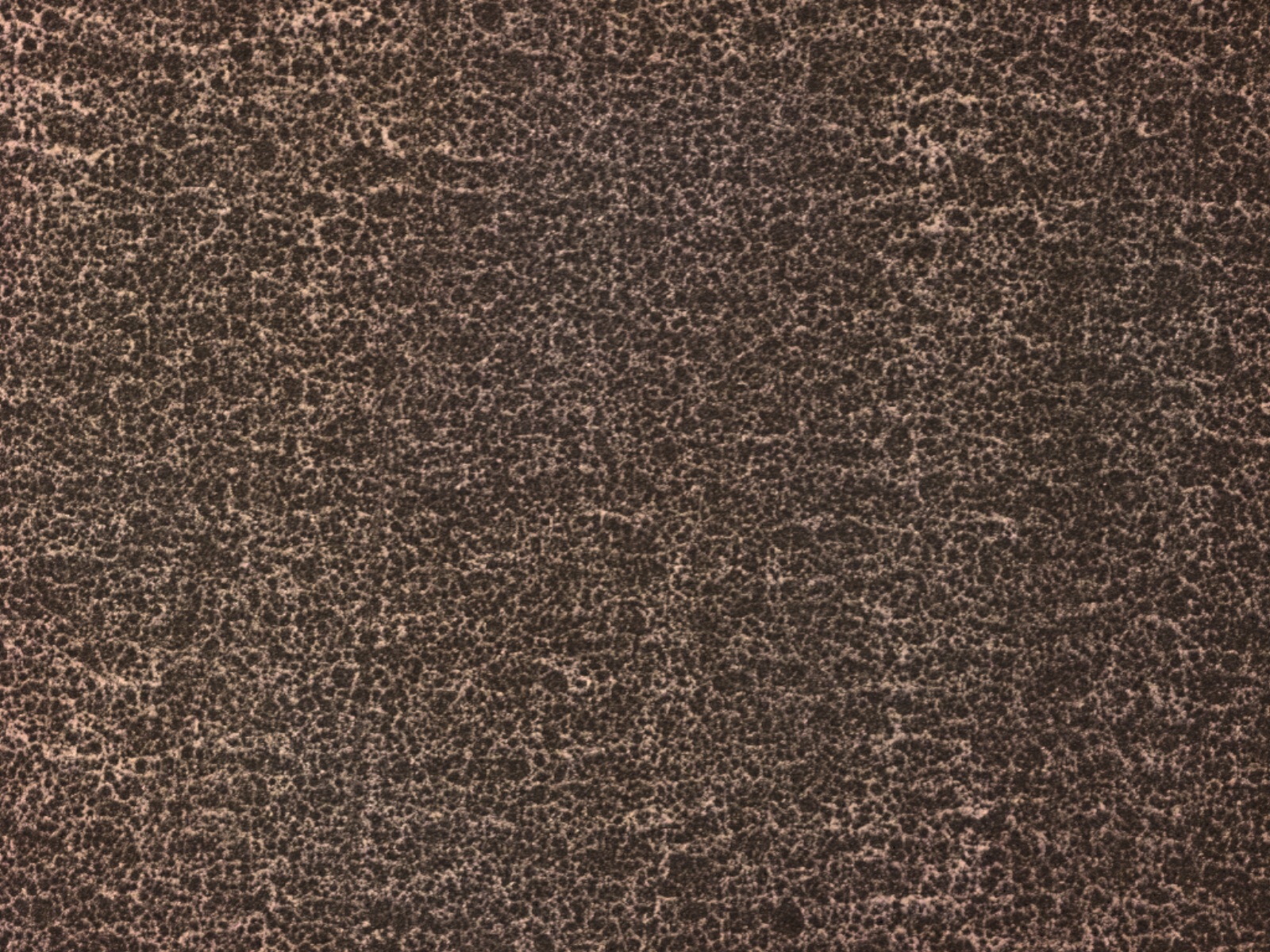}
    \includegraphics[width=0.4\textwidth, height = 0.4\textwidth]{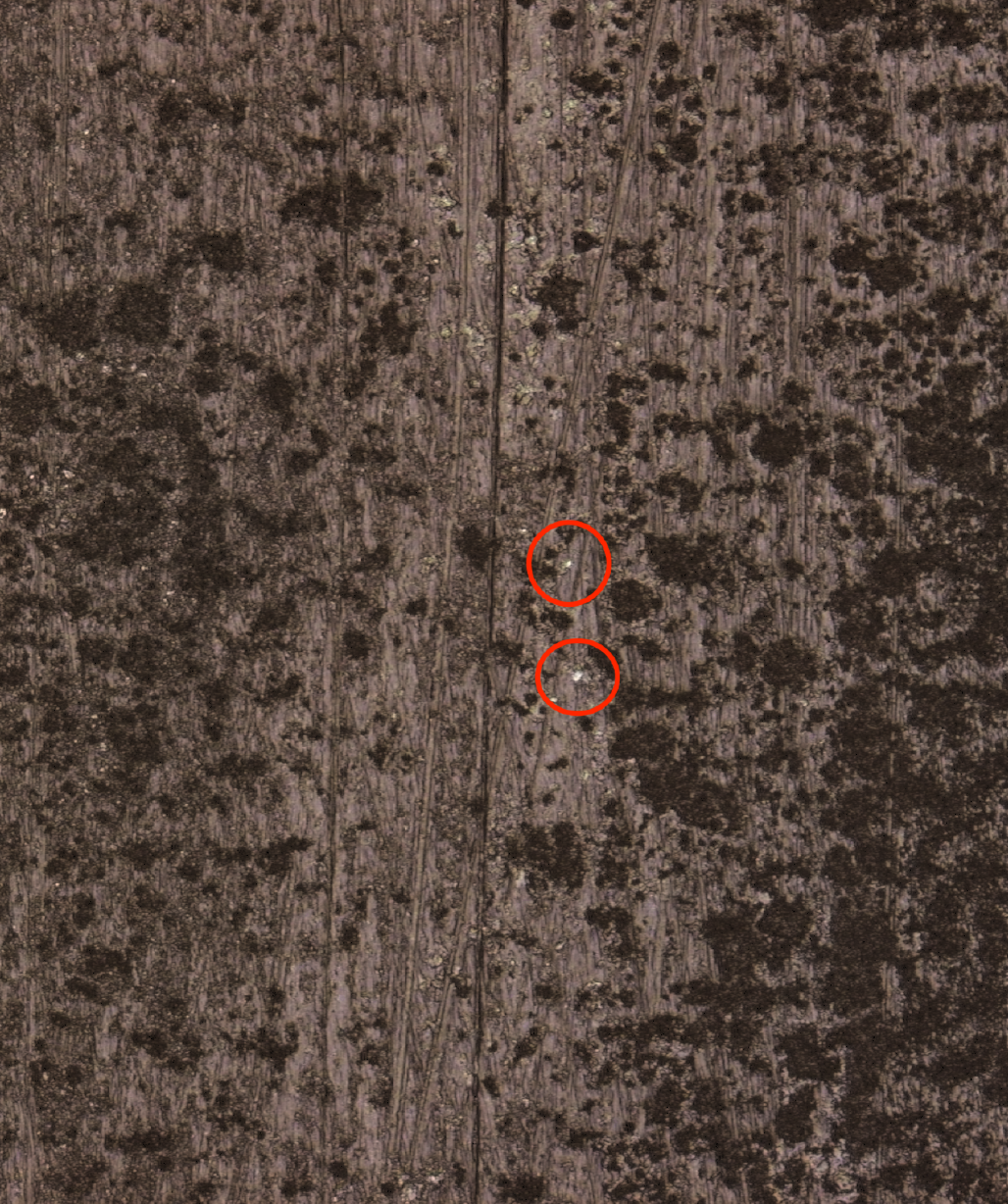}
   \caption{PTFE before plasma exposition; exposed plasma source (1) with sputtered gold particles (gold in the red circles)}
   \label{fig:PTFE}
\end{figure}
PTFE showed an interesting behavior. Since the effects of plasma exposure increased in XPS data from source (1) to source (3), the surface roughness decreased at (1), it remained constant at (2) and increased by a factor of 2.5 at (3). Figure \ref{fig:PTFE} showed the flatter areas of plasma source (1) after plasma exposure in the middle of the right image. These effects were much more pronounced when the areas of the filaments were analyzed. There the loss of material reached a depth of 26 - 66 µm for PTFE. This finding is in line with the findings concerning the sputtering of PTFE surfaces \cite{https://doi.org/10.1002/sia.740160186}. The stronger filaments from source (1) smoothed the surface, due to the sputtering of vertical structures and the lack of filamentary etched areas. Those were perpendicular to the down stream to the HV electrode. Since the filaments were smaller and the down-stream velocity is lower, the other sources did not face this smoothing effect. Deposits of gold on the dielectric were detectable with the microscope. These were seen as slightly bright dots in the centre of the image in figure \ref{fig:PTFE} (marked). This effect also increased with stronger filament formation and sputtering from source (3) to (1).
\subsubsection*{FR-4}
\begin{figure}[ht!]
   \centering
   \includegraphics[width=0.4\textwidth, height = 0.4\textwidth]{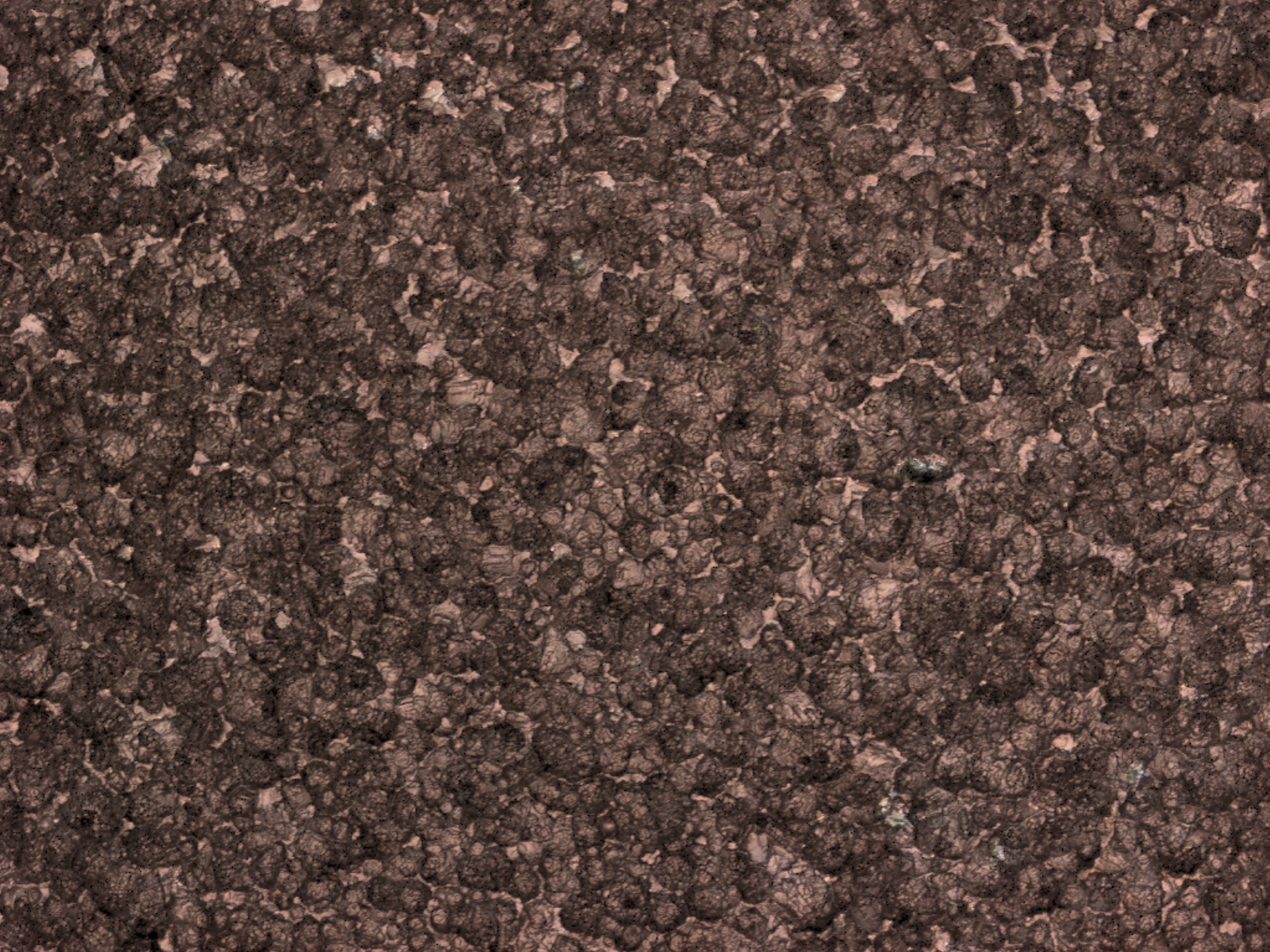}
   \includegraphics[width=0.4\textwidth, height = 0.4\textwidth]{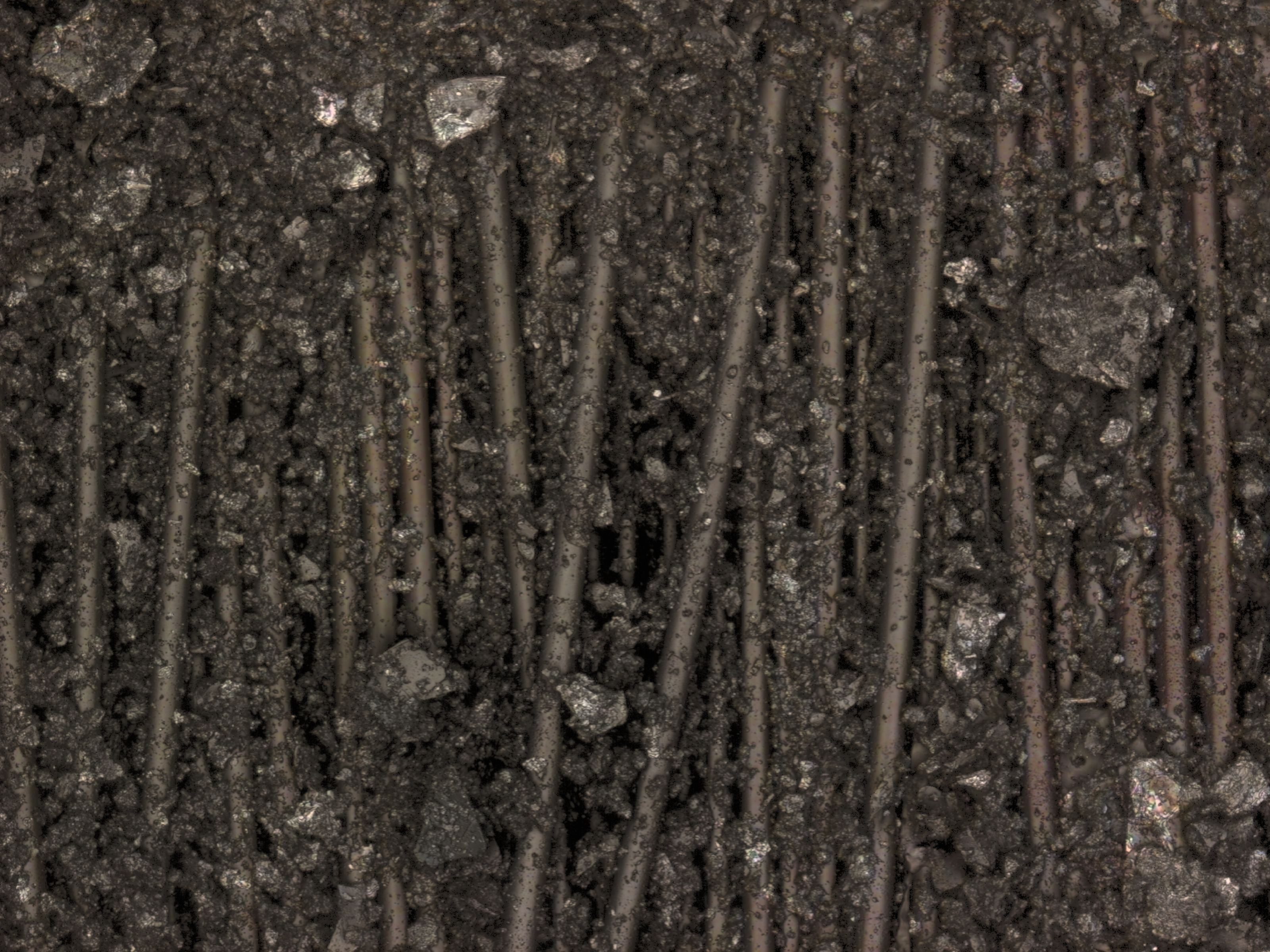}
   \caption{FR-4 surface before plasma exposition; plasma etched area with visible fiberglas inlay}
   \label{fig:Glasfaser}
\end{figure}
The most remarkable results were measured with the FR-4 dielectric, because it showed the lowest resistance to atmospheric plasma. The material changes were clearly visible in the microscope images. The roughness values also show significant variations. At 50x magnification, the surface roughness doubles from pre- to post-exposition. In addition, indentations in the material can be seen, which are formed at the exact positions of the plasmas filaments. The etched depth is 33 - 120µm, which means that the fiber-glass structure of the PCB material is exposed. The individual fibers can be clearly seen in figure \ref{fig:Glasfaser}. The fillers for the high frequency behavior of the FR-4 were also visible as larger particles.
These surface changes were comparable with the other plasma sources. The etched depth and the roughness changed depend on the strength of the filament formation. Therefore the effect decreases from plasma source (1) with the strongest filaments to (3) with a very homogenous plasma.

\subsubsection*{Electrodes}
\begin{figure}[ht!]
   \centering
\includegraphics[angle = 90, width=0.4\textwidth, height = 0.4\textwidth]{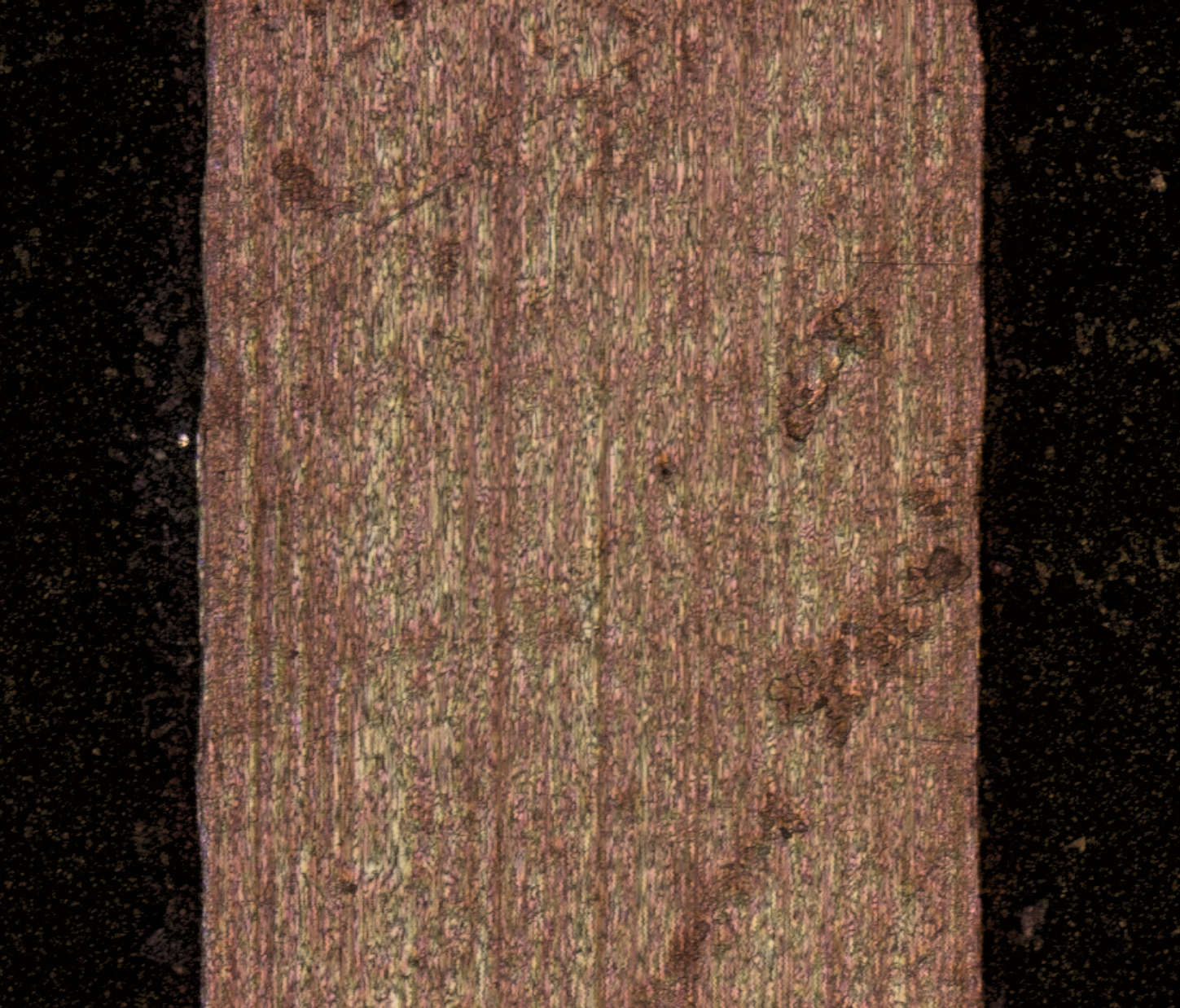}
\includegraphics[width=0.4\textwidth, height = 0.4\textwidth]{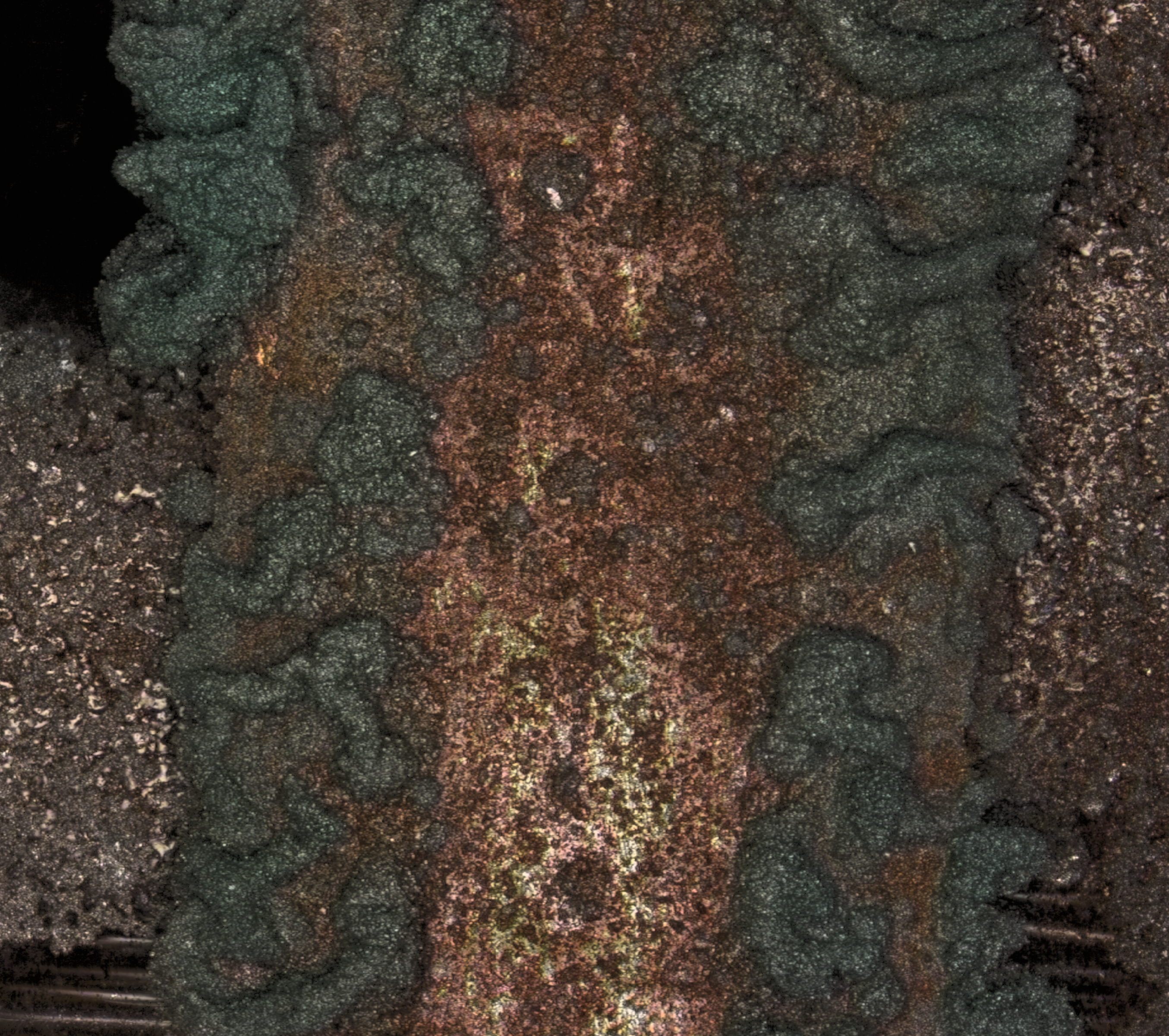}\\

\includegraphics[width=0.4\textwidth, height = 0.4\textwidth]{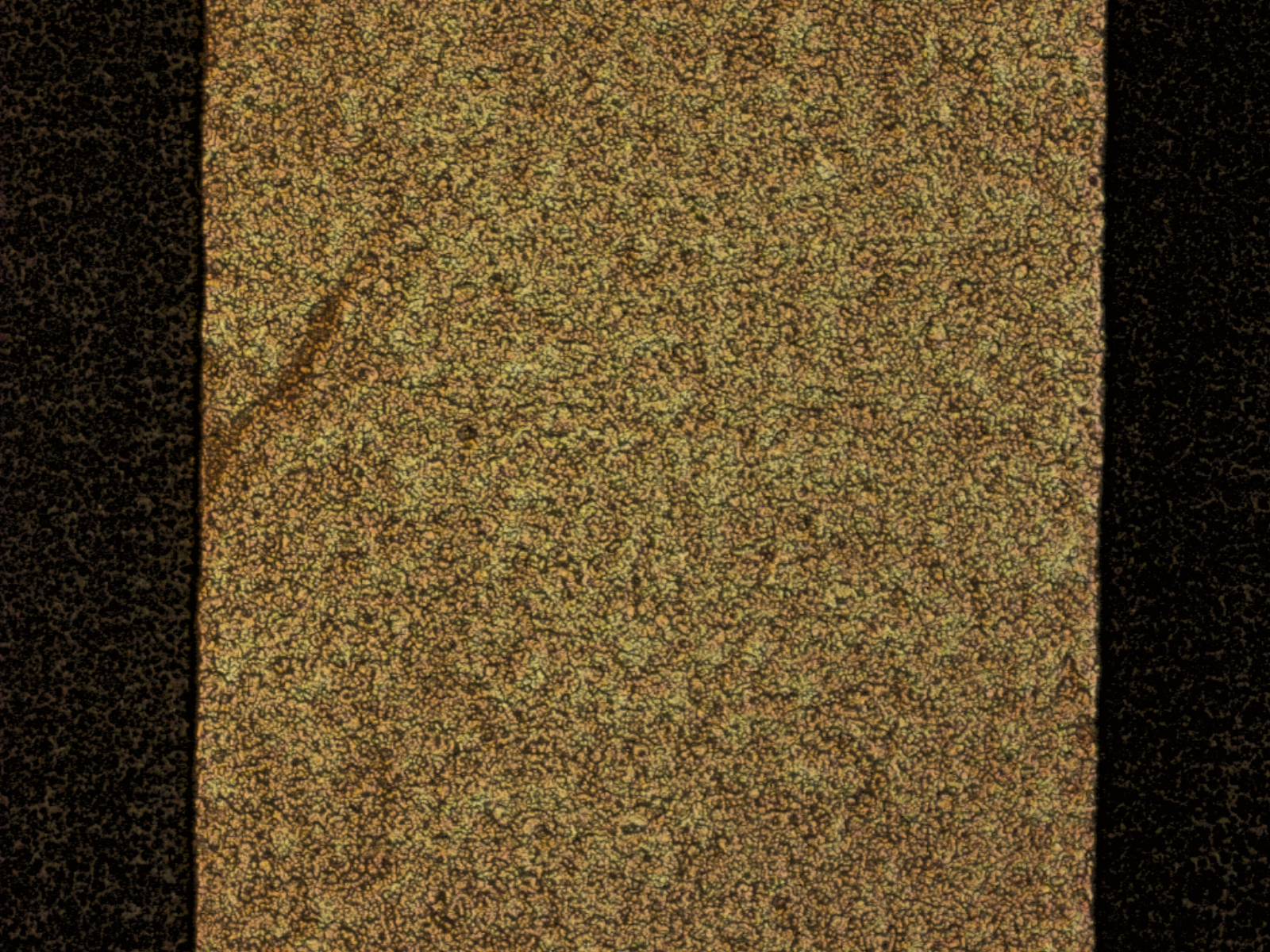}   
   \includegraphics[width=0.4\textwidth, height = 0.4\textwidth]{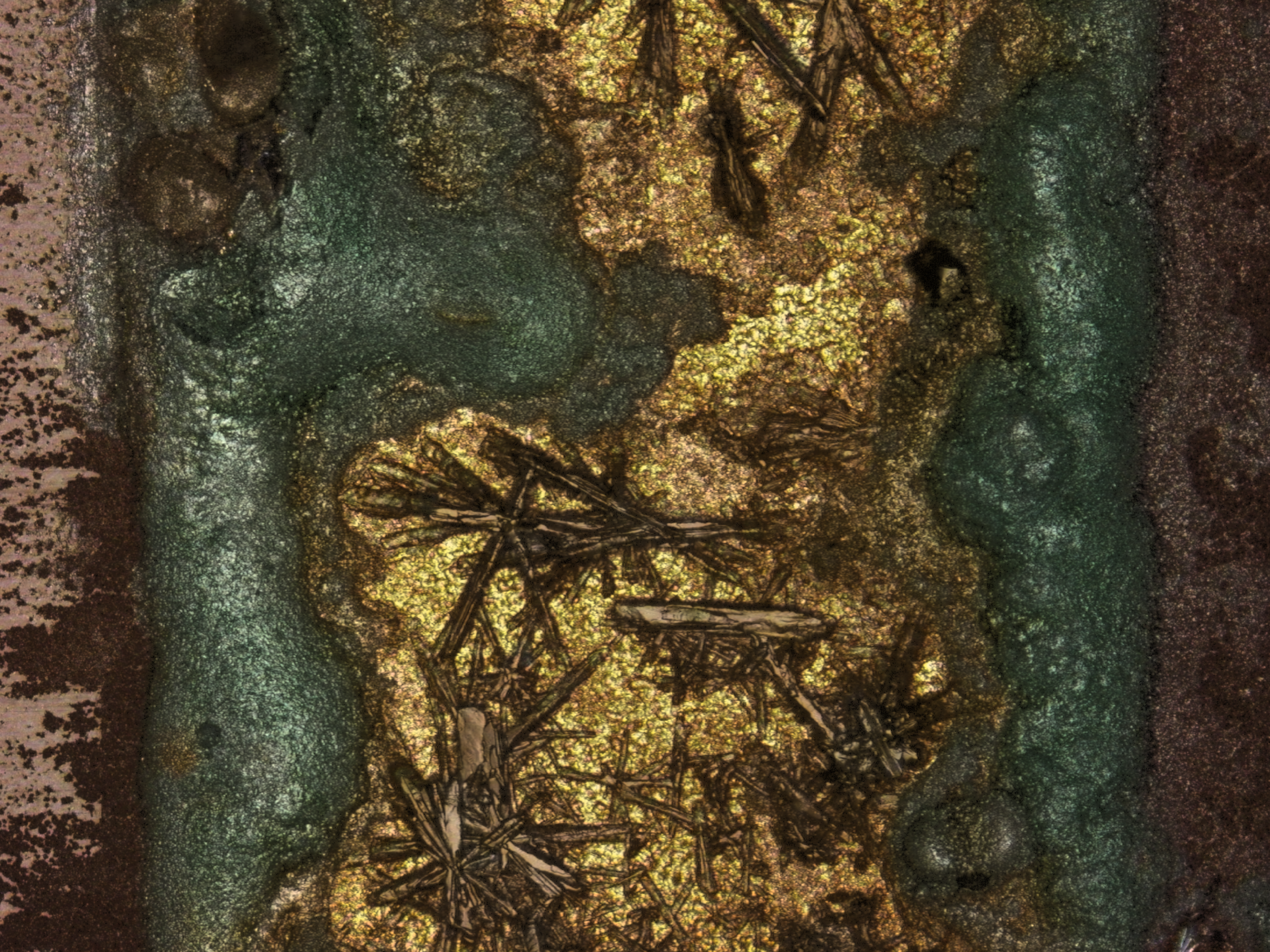}
   \caption{Electrodes before plasma exposure (left); formation of verdigris after plasma exposition (right); copper traces on FR-4 (up); gold plated copper traces on PTFE (down)}
   \label{fig:Leiterbahn}
\end{figure}
The roughness of the silver conductor tracks is significantly higher than that of the other plasma sources. This is due to the sintering process. Yet, neither the roughness nor the visible surface analysis was able to detect changes during plasma exposition. Therefore, those are not displayed in figure \ref{fig:Leiterbahn}. \\
The copper traces behaved as expected. The copper reacted with the carbon from the FR-4 and air humidity to verdigris, which is copper carbonate hydroxide. This was formed on the top of the entire conductor track. However, the formation of verdigris was increased in the area of the filaments.
The gold-coated conductor tracks were different. Due to the detectable gold particles in figure \ref{fig:PTFE}, the plasma sputtered the coating locally and formed verdigris at these spots. The result can be seen in right bottom part of figure \ref{fig:Leiterbahn}.\\
The changes of the three electrode materials (copper, silver, gold) under the plasma chemical treatment varied. The most interesting results were filamentary shaped deposits at the aluminium oxide surface as seen in figure \ref{fig:Al2O3nachher}. The rough surface is filled with the silver particles, which influences the roughness measurement. The gold plated electrodes of the PTFE were not resistant to plasma because the plasma etching capability of the plasma was the strongest within its filaments and removed the coating at the edges of the electrodes. Then the plasma oxidized the copper under the gold plating and the known behavior from FR-4 occurred.

\section{Conclusion}
 The experimental results, derived from both X-ray photoelectron spectroscopy (XPS) and laser microscopy, elucidate the varying degrees of surface and chemical modifications induced by plasma exposure across these materials. However, the higher permittivity of aluminum oxide leads to a higher displacement current, reactive power and thus to higher capacitive losses. Aluminium oxide exhibited remarkable chemical stability under plasma exposure, with no significant changes in the aluminum-to-oxygen ratio or the oxidation state of aluminum. This stability confirms its suitability for plasma source applications. It does not experience any significant change in surface roughness due to the plasma. Small variations can be seen due to the application of silver material in the recesses of the surface structure, resulting in a slightly lower surface roughness. This is caused by silver deposition, which can be observed locally in the areas of the filaments and decrease in intensity from plasma source (1) to (3), as does the strength of the filament formation and the roughness. The sputtered particles become negatively charged by the fast electrons and deposit above the dielectric covered high voltage electrode from plasma source (1) during the positive half cycle. \\
The PTFE underwent notable chemical alterations, primarily the oxidation of carbon chains and the substitution of fluorine atoms with oxygen. These changes, which include the formation of carbonyl, hydroxyl, and aldehyde groups, suggest a decrease in the material's inertness under long-term plasma exposure. It shows an interesting behavior. Since the effects of plasma exposure increase in XPS data from source (1) to (3), the surface roughness decreases at (1), it remains constant at (2) and increases by a factor of 2.5 at (3). Additionally, the stronger filaments from source (1) smoothens the surface, due to the sputtering of exhibited structures. During the positive half cycle of the high voltage, the negatively charged particles stream to the dielectric covered HV electrode. This results in deposits of gold on the dielectric. These can be seen as slightly bright dots in the centre of figure \ref{fig:PTFE}. This effect also increases with stronger filament formation from source (1) to (3).\\
The FR-4 showed the most substantial modifications among the tested materials, which is displayed in figure \ref{fig:Glasfaser}. The plasma exposure led to the etching of the epoxy matrix and the exposure of underlying fibre glass and filler, significantly altering the surface's physical and chemical properties. The increase in oxidized carbon content and the presence of nitrogen signals indicate a high reactivity to plasma, which compromises its structural integrity. FR-4 has the lowest resistance to atmospheric plasma. The material changes are clearly visible in the microscope images. Indentations in the material can be seen, which are formed at the exact positions of the filaments. The etched depth is 33 - 120 µm, which means that the fiber-glass structure of the PCB material is exposed. The individual fibers can be clearly seen. The fillers for the high frequency behavior of the FR-4 can also be seen.The etch depth and the roughness change depend on the strength of the filament formation. Therefore the effect decreases from plasma source (1) to (3).\\
The study also analyzed the effects of atmospheric plasma on three different electrode materials commonly used in plasma sources: copper, silver, and gold-plated copper. At first, the copper electrodes, used in conjunction with FR-4 dielectric material, reacted with carbon from FR-4 to form verdigris. This formation was more pronounced in areas exposed to plasma filaments, indicating a localized effect of the plasma on the electrode surface.\\
The silver electrodes were sintered onto aluminium oxide (Al\(_2\)O\(_3\)) dielectric material, resulting in a rough surface texture due to the sintering process. Plasma exposure led to a filamentary-shaped sputtering of the sintered silver particles, filling the recesses of the dielectric and influencing the overall roughness. It is suggested that this is caused by a sputtering process.\\
The gold-plated copper electrodes faced a sputtering of the coating. This effect was strongest within the filaments, removing the gold at the edges of the electrodes and distributing it over the surface. Once the gold plating was sputtered and the underlying copper was accessible, the plasma could oxidized those, leading to the same formation of verdigris, as the FR-4 showed.\\
The resilience of Al\(_2\)O\(_3\) against plasma-induced changes underscores its potential for use in environments where chemical and thermal stability are needed. The transformations observed in PTFE highlight the need for careful consideration of material specifications and potential protective measures in applications involving prolonged plasma exposure. The significant degradation of FR-4 suggests its limited utility in plasma environments. Due to the simple processing and compared to other results no significant changes of the plasma ignition behavior, shape of the filaments and plasma chemistry, it is a suitable dielectric for early stage development purposes and plasma design validation \cite{Joni}.\\
Further research is recommended to explore the long-term effects of plasma exposure on these materials over extended periods and under varying environmental conditions. Additionally, the development of protective coatings or modifications to the plasma exposure protocols could enhance the durability and performance of sensitive materials like PTFE and FR-4.\\
Further research for directly applied medical plasma sources should consider the combination of sintered gold electrodes on aluminum oxide. Due to the toxicology of the other materials and the lack of oxidation products of gold and alumina, those should be used for further developments.
\section{Acknowledgements}
We would like to express our gratitude to the Center for Materials Research at the Justus Liebig University Giessen for their support during XPS measurements. We also extend our sincere thanks to the German Aerospace Center (DLR) for providing funding for Alisa Schmidt's (grant \# 50EX2368) work on this project.

\bibliographystyle{unsrt}%
\bibliography{Literaturverzeichnis.bib}
\addcontentsline{toc}{section}{Literaturverzeichnis}
\end{document}